\documentclass[seceq]{ptptex}

\usepackage{graphicx}
\usepackage{bm}
\usepackage{wrapft}

\title{ A Dynamical System with Q-deformed Phase Space \\ Represented in Ordinary Variable Spaces}

\author{S. Naka$^{1}$, H. Toyoda$^{2}$ and T. Takanashi$^{1}$}

\inst{$^1$Department of Physics, College of Science and Technology
Nihon University,\\ 1-8-14 Kanda-Surugadai Chiyoda-ku Tokyo, Japan \\
$^2$Junior College, Funabashi Campus, Nihon University,\\ 7-24-1 Narashinodai Funabashi-shi Chiba, Japan }

\abst{
ABSTRACT: Dynamical systems associated with a q-deformed two dimensional phase space are studied as effective dynamical systems described by ordinary variables. In quantum theory, the momentum operator in such a deformed phase space becomes a difference operator instead of the differential operator. Then, using the path integral representation for such a dynamical system, we derive an effective short-time action, which contains interaction terms even for a free particle with q-deformed phase space. Analysis is also made on the eigenvalue problem for a particle with q-deformed phase space confined in a compact space. Under some boundary conditions of the compact space, there arises fairly different structures from $q=1$ case in the energy spectrum of the particle and in the corresponding eigenspace . }

\begin{document}

\maketitle

\section{Introduction}

One-dimensional dynamical system associated with the q-deformed phase space~\cite{q-phase-space,S-Wess,C-H-M-Wess} $px=qxp$ is known to lead to a quantum mechanical system having the canonical pairs characterized by a q-deformed Heisenberg algebra. The q-deformation, then, causes a drastic change of dynamical structure like a spontaneous symmetry breaking of the system. For example, in such a system, the meaning of translational invariance is changed even for a free particle\cite{Representation-q-H.A.}~; and there arises a new type of discrete symmetry~\cite{lattice-structure}~. 

As an application of the q-deformation for the particle physics, it may be interesting to consider a particle embedded in a higher-dimensional spacetime with deformed extra dimensions. The deformation, then, breaks a higher-dimensional symmetry and yields a fairly different structure of excitation spectrum from $q=1$ case. When we study such a deformed extra dimension, it is convenient to use a coordinate representation of the q-deformed Heisenberg algebra constructed out of ordinary variables. This is due to the reason that compactness conditions for the extra dimensions are naturally applied not to the variables in non-commutative spaces but to those in ordinary spaces. 

According to this point of view, it is worthwhile to study the representation of the q-deformed Heisenberg algebra~\cite{W-Zumino}
\begin{equation} 
 \hat{\partial}_x x-qx\hat{\partial}_x=1  \label{q-derivative}
\end{equation}
 coming from q-deformed phase space by a combination of the $x$ and its ordinary derivative operator $\partial_x=\partial/\partial x$ in such a way that
\begin{equation}
 \hat{\partial}_x=\partial_x\frac{q^{(x\partial_x)}-1}{(q-1)(x\partial_x)} . \label{hated derivative}
\end{equation}
This equation defines a mapping between the space of deformed calculus and that of ordinary calculus; as a result of this mapping, the momentum operator in q-deformed quantum theory becomes a difference operator instead of the differential operator. The purpose of this paper is, thus, to study the particle dynamics in  q-deformed phase space as an effective theory of those in ordinary phase space through the above mapping.  

In the next section, we discuss a free particle in two-dimensional q-deformed phase space, which leads to one-dimensional Schr\"odinger equation for the particle without any specific boundary conditions in quantum mechanics. Then, using a path integral representation for such a dynamical system, we derive an effective short-time action, which contains interaction terms in ordinary phase space even for a free particle in the q-deformed phase space. 

Section 3 is the discussion on the eigenvalue problem for the q-deformed particle confined in a compact space; there, we confine our attention to the case of a particle in infinite potential well $0<x<L$. There, the eigenvalue problem of the Hamiltonian under this situation is discussed in detail. We, then, show the fairly different structure from $q=1$ case in the energy spectrum of the particle and the corresponding eigenspace.

Section 4 is the discussion and summary. In appendix A and B, the discussion is made on the mathematical background of sections 2 and 3.

\section{Free particle associated with q-deformed phase space}

In the usual phase space dynamics, the translation from classical theories to quantum mechanical counterparts can be carried out by the substitution $p\rightarrow -i\hbar\partial_x$ in the Hamiltonian operator of a dynamical system. However, since the operator (\ref{hated derivative}) is not an anti-hermitian operator, $-i\hbar\hat{\partial}_x$ is not a hermitian momentum operator. Then, it is necessary to modify the relation between $\hat{\partial}_x$ and the momentum operator so that 
\begin{equation}
 \hat{p}_q =-\frac{i\hbar}{2}(\hat{\partial}_x+\hat{\partial}_x^\dag)
         =-i\hbar\left(\frac{q+1}{2q}\right)D_x~, \label{q-momentum}
\end{equation}
to get a hermitian momentum operator $\hat{p}_q$. Here, $D_x$ can be found to be 
\begin{equation}
  D_x=x^{-1}[\hat{N}]~,~~\left( ~[a]=\frac{q^a-q^{-a}}{q-q^{-1}}~,~~\hat{N}=x\partial_x~ \right), \label{difference-1}
\end{equation}
and it satisfies
\footnote{
To realize the algebra (\ref{q-derivative}), we may change the role of $\hat{\partial}_x$ and $x$. Namely, the ordinary derivative $\partial_x$ and the deformed coordinate $\hat{x}=\frac{q^{(x\partial_x)}-1}{(q-1)(x\partial_x)}x$ form another set of operators satisfying (\ref{q-derivative}). In this case, since $\hat{x}$ is not a hermitian operator, the operator $X=\frac{1}{2}(\hat{x}+\hat{x}^\dag)=\frac{q+1}{2q}\frac{[x\partial_x]}{x\partial_x}x$ becomes a physical coordinate operator, to which one can verify $\partial_x X-qX\partial_x=\frac{q+1}{2q}q^{-x\partial_x}$. Substituting, here,  $-i\hbar\partial_x$ for $\hat{p}$, the q-commutator (\ref{q-canonical}) with $q^{-x\partial_x}$ factor in the right-hand side is again obtained.

We further note that the operators $x^2,D^2$ and $\hat{N}+\frac{1}{2}$ form a $SU_q(2)$ like generators defined by 
\[ [\hat{N}+\frac{1}{2},x^2]=2x^2,~[\hat{N}+\frac{1}{2},D^2]=-2D^2,~{\rm and}~[D^2,x^2]=[2(\hat{N}+\frac{1}{2})]. \]
}
\begin{equation}
 D_x x-q^{\pm 1}xD_x=q^{\mp \hat{N}}. \label{difference-2}
\end{equation}
The operator $D_x$ has the meaning of q-difference acting on a function of $x$ as
\begin{equation}
 D_xf(x)=\frac{f(qx)-f(q^{-1}x)}{(q-q^{-1})x} ~, \label{difference-3}
\end{equation} 
which will reduce to the usual differential operator $\partial_x$ as $q\rightarrow 1$. From equations (\ref{q-momentum}) and (\ref{difference-2}), the q-deformed momentum operator $\hat{p}_q$ satisfies
\begin{equation}
 \hat{p}_q x-q^{\pm 1}x\hat{p}_q=-i\hbar\left(\frac{q+1}{2q}\right) q^{\mp \hat{N}} ~, \label{q-canonical}
\end{equation}
which is equivalent to the following usual commutation relation:
\begin{equation}
 [\hat{p}_q,x]=-\frac{i\hbar}{2}\left(q^{\hat{N}}+(q^{\hat{N}})^\dag \right)~.
\end{equation}
Then, under a state $\psi(x)$ with $\langle \psi|\psi\rangle=1$, one can obtain the uncertainty relation between $p_q$ and $x$ such as
\begin{equation}
 \Delta p_q \Delta x \geq \frac{\hbar}{4}\left|\int_{-\infty}^\infty dx\left(\psi(x)^*\psi(qx)+\psi^*(qx)\psi(x) \right)\right|~.
\end{equation}

The Sch\"rodinger equation based on this deformed calculus is
\begin{equation}
 i\hbar\frac{\partial}{\partial t}\psi(t,x)=\hat{H}_q\psi(t,x) ~.
\end{equation} 
Here $\hat{H}_q$ is the Hamiltonian operator, in which the momentum operator is given by (\ref{q-momentum}). For a particle of mass $m$, the $\hat{H}_q$ becomes
\begin{equation}
 \hat{H}_q =\frac{1}{2m}\hat{p}_q^2+V(x)=-\frac{\hbar^2}{2m_q}D_x^2+V(x) ~,
\end{equation}
where $ m_q=\left(\frac{q+1}{2q}\right)^2 m$. It should be noticed that the free Hamiltonian operator $\hat{H}^0_q=\frac{1}{2m}\hat{p}_q^2$ in the deformed calculus does not represent a free particle one in the ordinary calculus, since it can be written as
\begin{equation}
 \hat{H}^0_q=-\frac{\hbar^2}{2m_q}\frac{1}{(q-q^{-1})^2}x^{-1}\left[2\cos\left(\frac{1}{\hbar}\{x,\hat{p}\} \log q \right)-(q+q^{-1})\right]x^{-1}~, \label{quantum-H0}
\end{equation}
where $\hat{p}=-i\hbar\partial_x$ is the momentum operator in the ordinary calculus. The Hamiltonian operator (\ref{quantum-H0}), however, is not a classical Hamiltonian itself, since the form is depending on the operator ordering. As the next task, thus, we try to derive a classical counterpart of $\hat{H}_q^0$ in ordinary calculus by means of the path integral method
\footnote{
There is another approach to the path integral representation to the propagator~\cite{path-integral} based on the q-deformed phase space. Since the configuration space is different from ordinary space, the result is different from ours.}. For this purpose, the role of the potential $V(x)$ is not important; and so, we only consider the $V=0$ case in what follows.

To study the propagation kernel by the free Hamiltonian $\hat{H}_q^0$ for the finite time interval $T=t-t^\prime$, let us introduce the eigenstate of $\hat{p}_q$ such that
\begin{equation}
 \hat{p}_q|k,q\rangle=\hbar k|k,q\rangle~~~{\rm and}~~~\langle k,q|k^\prime,q\rangle=\delta(k-k^\prime)~.
\end{equation} 
The $x$-representation of this state is given explicitly by the q-exponential function $\langle x|k,q\rangle=\frac{1}{\sqrt{2\pi\Gamma_q[1]}}e_q^{ikx}$, which tends to the usual plane wave function $\langle x|k\rangle=\frac{1}{\sqrt{2\pi}}e^{ikx}$ as $q\rightarrow 1$ (Appendix A) .

In terms of this eigenstate, the propagation kernel can be written as
\begin{equation}
 \langle x|e^{-\frac{i}{\hbar}T\hat{H}_q^0}|x^\prime\rangle =\int_{-\infty}^\infty dk\langle x|k,q\rangle e^{-\frac{i}{\hbar}T\frac{(\hbar k)^2}{2m_q}}\langle k,q|x^\prime\rangle~. \label{finite-T}
\end{equation}
We note that the $\langle x|k,q\rangle\langle k,q|x^\prime\rangle$ is not a function of $x-x^\prime$, and so, the translational invariance of this kernel is lost due to $q\neq 1$.

Now, according to (\ref{q-exponential-product}), the $\langle x|k,q\rangle\langle k,q|x^\prime\rangle$ can be expressed as
\begin{equation}
 \langle x|k,q\rangle\langle k,q|x^\prime\rangle=\frac{1}{2\pi\Gamma_q[1]}\sum_{N=0}^\infty \frac{(ik)^N}{[N]!}(x\dot{-}x^\prime)_q^N ~.
\end{equation}
Substituting this equation for (\ref{finite-T}), and carrying out the Gaussian integral with respect to $k$, we obtain the series out of $N=2n$ terms such that
\begin{equation}
 \langle x|e^{-\frac{i}{\hbar}T\hat{H}_q^0}|x^\prime\rangle=\frac{1}{\Gamma_q[1]}\sqrt{\frac{m_q}{2\pi i\hbar T}}\sum_{n=0}^\infty\frac{(2n)!}{n![2n]!}\left(\frac{i}{\hbar}\frac{m_q}{2}\right)^n\frac{(x\dot{-}x^\prime)_q^{2n}}{T^n} ~. \label{q-propagation-kernel}
\end{equation}
This right-hand side of this equation gives the exact form of propagation kernel in a finite time interval $T$, which is reduced to the ordinary free propagation kernel $\langle x|e^{-\frac{i}{\hbar}T\hat{H}^0}|x^\prime\rangle=\sqrt{\frac{m}{2\pi i\hbar T}}\exp\{\frac{i}{\hbar}\frac{m}{2}\frac{(x-x^\prime)^2}{T}\}$ in the limit $q\rightarrow 1$. However, since the q-binomial expansions $(x\dot{-}x^\prime)_q^{2n},(n=1,2,\cdots)$ have common zero points at $x=q^{\pm}x^\prime$ by (\ref{product-formula}) instead of $x=x^\prime$, the q-propagation kernel becomes translational invariant only for the interval $(x,q^{\pm}x)$; that is, we obtain $\langle q^{\pm}x|e^{-\frac{i}{\hbar}T\hat{H}_q^0}|x\rangle=\frac{1}{\Gamma_q[1]}\sqrt{\frac{m_q}{2\pi i\hbar T}}$, which should be compared with $\langle x|e^{-\frac{i}{\hbar}T\hat{H}^0}|x \rangle=\sqrt{\frac{m}{2\pi i\hbar T}}$ in the ordinary free propagation kernel.

The transition amplitude between a finite time interval $T$ by the Hamiltonian $\hat{H}_q^0$ can be written as 
\begin{equation}
 \langle x_b|e^{-\frac{i}{\hbar}T\hat{H}_q^0}|x_a\rangle=\lim_{N\rightarrow \infty}\int \left(\prod_{i=1}^{N-1}dx_i \right) \prod_{i=1}^N \langle x_i|e^{-\frac{i}{\hbar}\Delta t\hat{H}_q^0}|x_{i-1} \rangle ~,
\end{equation}
where $x_N=x_b$, $x_0=x_a$, and $\Delta t=T/N$. The classical action associated with $\hat{H}_q^0$ is, then, appears as the phase factor of the transition kernel between two neighboring points $(x_i,x_{i-1})$ with short time interval $\Delta t$. Within the first order approximation of $\Delta t$, we can evaluate the kernel as
\begin{equation}
 \langle x_i|e^{-\frac{i}{\hbar}\Delta t\hat{H}_q^0}|x_{i-1} \rangle \simeq \langle x_i|x_{i-1}\rangle + \frac{i}{\hbar}\Delta t\frac{\hbar^2}{2m_q}D_i^2\langle x_i|x_{i-1}\rangle~,  \label{short-time-kernel}
\end{equation}
which is consistent with (\ref{finite-T}) within the approximation up to the first order of $T$, since $\langle x|k,q\rangle k^2 \langle k,q|x^\prime\rangle=-D_x^2\langle x|k,q\rangle\langle k,q|x^\prime\rangle$ and $\int_{-\infty}^\infty dk\langle x|k,q\rangle\langle k,q|x^\prime\rangle=\langle x|x^\prime\rangle$.

Using the momentum expansion of $\langle x_i|x_{i-1}\rangle=\int_{-\infty}^\infty\frac{dp}{2\pi\hbar}e^{ip(x_i-x_{i-1})}$, the second derivative of $\langle x_i|x_{i-1}\rangle$ by $D_i$ in the right-hand side can be evaluated as
\begin{align}
D_i^2\langle x_i|x_{i-1}\rangle & =\int\frac{dp}{2\pi\hbar}\frac{q^{-1}e^{ip(q^2-1)x_i/\hbar} +qe^{ip(q^{-2}-1)x_i/\hbar}-(q+q^{-1})e^{ip(x_i-x_{i-1})/\hbar} }{(q-q^{-1})^2x_i^2}  \nonumber \\
 & =\int\frac{dp}{2\pi\hbar}e^{ip(x_i-x_{i-1})/\hbar}\frac{q+q^{-1}}{(q-q^{-1})^2\left(\frac{x_i+x_{i-1}}{2}\right)^2} \left\{ \cos\left( p\frac{q-q^{-1}}{q+q^{-1}}\frac{x_i+x_{i-1}}{\hbar} \right)-1 \right\} \label{second-derivative}
\end{align}
Substituting (\ref{second-derivative}) for (\ref{short-time-kernel}), the short-time-propagation kernel can be written as $\langle x_i|e^{-\frac{i}{\hbar}\Delta t\hat{H}_q^0}|x_{i-1} \rangle \simeq \int\frac{dp}{2\pi\hbar}\exp\{\frac{i}{\hbar}\Delta tL(p,x_i,x_{i-1})\}$. Here, $L(p,x_i,x_{i-1})$ is the phase space Lagrangian of the system in the interval $(x_i,x_{i-1})$, which formally tends to
\begin{equation}
 L(p,x,\dot{x})=p\dot{x}-\frac{\hbar^2}{2m_q}\frac{q+q^{-1}}{(q-q^{-1})^2x^2} \left\{1- \cos\left( 2p\frac{q-q^{-1}}{q+q^{-1}}\frac{x}{\hbar} \right) \right\}
 \label{p-Lagrangian}
\end{equation}
by assuming $x_i-x_{i-1}\sim 0(\Delta t)$ according as $\Delta t\rightarrow 0$. The variable $p$ has the meaning of momentum conjugate to $x$ because of $p=\frac{\partial L}{\partial \dot{x}}$; and so, the Hamiltonian of the system $H(x,p)=p\dot{x}-L$ becomes
\begin{equation}
 H(x,p)=\frac{\hbar^2}{2m_q}\frac{q+q^{-1}}{(q-q^{-1})^2x^2}2\sin^2\left( p\frac{q-q^{-1}}{q+q^{-1}}\frac{x}{\hbar} \right)~,
\end{equation}
to which one can verify $H(x,p)\rightarrow \frac{1}{2m}p^2,~(q\rightarrow 1)$. If we put $H=E(=const.)$, we obtain trajectories in the $(x,p)$ phase space, on which the total energy of the system is fixed to $E$.
\begin{figure}
\centerline{\includegraphics[width=4cm]{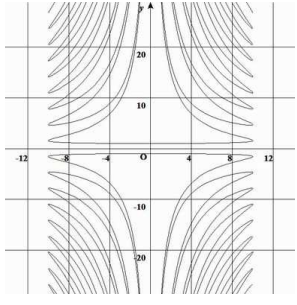}}
\caption{An equi-energy trajectory in phase space}
\label{fig:1}
\end{figure}
The trajectory is a straight line with $p_0=\sqrt{(q+q^{-1})m_qE}$ near $x=0$; then, the line will arrive at a turning point $x_{max}$ after stretching to some length (Fig.\ref{fig:1}), since $|x|$ is bounded by $|x|\leq \frac{\hbar}{|q-q^{-1}|}\sqrt{\frac{q+q^{-1}}{m_q E}}$ for a given $E$ 
\footnote{
On a $E$ fixed trajectory, the condition $dE=0$ leads to
\[ \frac{dx}{dp}=\frac{x^2}{\hbar}\frac{q-q^{-1}}{q+q^{-1}}\frac{1}{\tan(y)-y},~\left( y=p\frac{q-q^{-1}}{q+q^{-1}}\frac{x}{\hbar} \right). \]
Thus, $\frac{dx}{dp}=0$ is satisfied at $y=\frac{\pi}{2}n,(n=\pm 1,\pm 2,\cdots)$ in addition to $y=\infty$. The first maximal point $(x_1,p_1)$ in $x>0$ side is determined by $p_1\frac{q-q^{-1}}{q+q^{-1}}\frac{x_1}{\hbar}= \frac{\pi}{2}$ provided $q>1$; that is, we have $x_1= \frac{\pi\hbar}{2p_1}\frac{q+q^{-1}}{q-q^{-1}}$, which coincides with $x_{max}$ for $p_1=\frac{\pi}{2}p_0$. The sequential maximal points are obtained similarly at $(x_{max},\frac{n\pi}{2}p_0),(n=1,2,\cdots)$. }.

 Furthermore, the constraint $\frac{\partial L}{\partial p}=0$ allows us to solve $p$ as a function of $\dot{x}$, and we obtain
\begin{equation}
 p_n=\frac{\hbar}{2x}\frac{q+q^{-1}}{q-q^{-1}}\left[\sin^{-1}\left\{\frac{m_q}{\hbar}(q-q^{-1})x\dot{x}\right\}+2\pi n\right]~,~~(n=0,\pm 1,\pm 2, \cdots). \label{x-dot}
\end{equation}
In the above equation, $n$ implies the number that distinguishes branches of $\sin^{-1}\theta$ for $-1<\theta<1$. Substituting (\ref{x-dot}) for (\ref{p-Lagrangian}), the Lagrangian as a function of $x$ and $\dot{x}$ in the $n$-th branch is given by
\begin{align}
 L_n(x,\dot{x}) &=\frac{\hbar\dot{x}}{2x}\frac{q+q^{-1}}{q-q^{-1}}\left[\sin^{-1}\left\{\frac{m_q}{\hbar}(q-q^{-1})x\dot{x}\right\}+2\pi n\right] \nonumber \\
 &+\frac{\hbar^2}{2m_q}\frac{q+q^{-1}}{(q-q^{-1})^2x^2}\left[\sqrt{ 1-\left\{\frac{m_q}{\hbar}(q-q^{-1})x\dot{x}\right\}^2 }-1 \right]. \label{effective-action}
\end{align}
From this Lagrangian, by taking into account that $p\dot{x}$ term does not contribute to the Lagrange's equation, a little calculation leads to the equation of motion of $x$, for example for $n$=0, such that
\begin{equation}
 m_q\ddot{x}=-m_q\frac{\dot{x}^2}{x}-\frac{2\hbar^2}{m_q}\frac{\sqrt{1-\left\{\frac{m_q}{\hbar}(q-q^{-1})x\dot{x}\right\}^2}}{(q-q^{-1})^2 x^3} \left[ \sqrt{1-\left\{\frac{m_q}{\hbar}(q-q^{-1})x\dot{x}\right\}^2}-1 \right]~, \label{eq-of-motion}
\end{equation}
to which one can verify $\ddot{x}\rightarrow 0,~(q\rightarrow 1)$. It is interesting that the customary function for a harmonic oscillator $x(t)=A\sin\{\omega(t-t_0)\}$ becomes a solution of the above equation of motion provided that $\frac{m_q\omega A^2}{2}=\frac{\hbar}{|q-q^{-1}|}$
\footnote{The equation of motion (\ref{eq-of-motion}) can also be written as
\[ \frac{x}{\dot{x}}\frac{d}{dt}(\sqrt{1-y^2}-1)=2(\sqrt{1-y^2}-1),~\left(~y=\frac{m_q}{\hbar}(q-q^{-1})(x\dot{x})~\right). \]
Taking $\frac{x}{\dot{x}}\frac{d}{dt}=\frac{d}{d\log x}$ into account, this equation can be integrated easily to give $\dot{x}^2=-\left\{\frac{\hbar}{m_q(q-q^{-1})}\right\}^2(2C+C^2x^2)$ with a integration constant $C$. Choosing, here, $C=-\frac{2}{A^2}$ with the constraint in the main text, the \lq\lq $\sin$ \rq\rq function becomes a special solution of Eq.(\ref{eq-of-motion}).
}. This looks like that the free particle in q-deformed phase space allows bounded motions in ordinary phase space. The result, however, depends on the validity of the Lagrangian (\ref{effective-action}) beyond a short time interval.

\section{Particle in a box}

The particle in a box, a square well potential with perfectly rigid walls,  extending from $0$ to $L$,  is another interesting solvable problem\cite{Sol-Schrodinger-eq}, to which the results suffer fairly large modification from those in $q=1$ case. As usual, the eigenvalue problem of $\hat{H}_0$ is set as
\begin{equation}
 \hat{H}_0\psi(x)=-\frac{\hbar^2}{2m_q}D_x^2\psi(x)=E_q\psi(x),~(~0<x<L~) \label{box-1}
\end{equation}
with the boundary conditions
\begin{equation}
 \psi(0)=\psi(L)=0. \label{box-2}
\end{equation}
If we exclude singular solutions, the functional space of eigenstates are constructed out of a pair of q-deformed $\sin$ functions defined by
\begin{align}
 \sin_q(x) &= \sum_{n=0}^\infty \frac{(-1)^n}{[2n+1]!}x^{2n+1}~, \\
 {\rm\bar{s}in}_q(x) &= \sum_{n=0}^\infty \frac{(-1)^n}{[2n+1]!}q^{-\frac{1}{2}n(n-1)}x^{2n+1}~.
\end{align}
In what follows, we write simply $S_n(x)=N_n\sin_q(k_n x)$ and $\bar{S}_n(x)=N_n{\rm\bar{s}in}_q(k_n x)$, where $N_n$ is a normalization constant. By definition, these functions satisfy $S_n(0)=\bar{S}_n(0)=0$ obviously. Furthermore, one can show the followings  (Appendix B):\\

\begin{description}
\item[1)] The $\{S_n(x)\}$ are eigenfunctions of $D_x^2$ characterized by $D_x^2 S_n(x)=-k_n^2 S_n(x)$ and the orthogonality $\langle S_nS_m\rangle=\delta_{n,m}$. Here, $\langle \cdots \rangle$ is the $q$-integral defined by (\ref{q-integral}) with $a=0$ and $b=q^{-1}L$. Contrary to the case of $q=1$, however, the boundary condition $\psi(L)=0$ does not choose each of $\{S_n(x)\}$ directly.

\item[2)] The independent functions chosen by the boundary condition $\psi(L)=0$ are $\{\bar{S}_n(x)\}$, for which $k_n$'s are given by
\begin{equation} 
 k_n=\frac{\pi_q(n)}{L}~,~(n=1,2,\cdots),
\end{equation}
where $\pi_q(n)$ is the solution of $\pi n=\sum_{k=0}^\infty \tan^{-1}\{(1-q^{-2})q^{-2k}x\}$ with respect to $x$. Although, $\pi_q(x)$ is not an elementary function of $x$, it obviously tends to $\pi n$ in the limit $q\rightarrow 1$. The $\pi_q(n)$ can be expanded in an odd power series of $\pi n$, and its few terms are given by (\ref{b-1})-(\ref{b-7})
\begin{align}
 \pi_q(n) &=(\pi n)+\frac{1}{3}\frac{(1-q^{-2})^3}{1-q^{-6}}(\pi n)^3 \nonumber \\
 &+\left[-\frac{1}{5}\frac{(1-q^{-2})^5}{1-q^{-10}}+\frac{1}{3}\left\{\frac{(1-q^{-2})^3}{1-q^{-6}}\right\}^2 \right] (\pi n)^5 + \cdots
\end{align}

We note that each $\bar{S}_n(x)$ is not an eigenfunction of $D_x^2$ in the usual sense, since they satisfy $D_x^2\bar{S}_n(x)=-q^{-1}k_n^2\bar{S}_n(q^{-2}x)$; however, it can be verified that $\bar{D}_x^2\bar{S}_n(x)=-k_n^2\bar{S}_n(x)$ for $\bar{D}_x\equiv x^{-1}[\hat{N},q^2]$. 

\item[3)] Each $S_n(x)$ can be expanded as a linear combination of $\{\bar{S}_m(x)\}$ and vice versa; and, in this sense, the boundary condition at $x=L$ is also satisfied by $\{S_n(x)\}$.
\end{description} \vspace{3mm}

Therefore, the energy eigenvalues for a particle in the box under consideration can be written as
\begin{equation}
 E_n=\frac{\hbar^2}{2m_q}\left(\frac{\pi_q(n)}{L}\right)^2,~(n=1,2,\cdots).
\end{equation}
It should be, then, stressed that for a large deformation such as $q \geq 1.2$, the $\pi_q(n)$ increases rapidly in response to an increase in $n$, as can be seen from Table.\ref{table:1} and Fig.2. The result shows that $\log\pi_q(n)$ is almost linear in this interval of $n$. 

\begin{table}
\caption{Several approximate values of $\pi_q(n)$ for $q=1.5$.}
\label{table:1}
\begin{center}
\begin{tabular}{|c|c|c|c|c|} \hline
n & 1 & 2 & 3 & 4 \\ \hline
$\pi_q(n)$ & $5.28$ & $3.00\times 10$ & $1.55\times 10^2$ & $7.87\times 10^2$ \\ \hline
\end{tabular} \vspace{1mm}\\
\end{center}
\end{table} 

In this section, we have studied the eigenstates characterized by (\ref{box-1}) and (\ref{box-2}) within the framework of regular functions. If we allow a singular structure of wave functions, however, we may multiply $\{S_n(x)\}$ by the singular phase functions
\begin{equation}
 Q_m(x)=\exp\left\{i\frac{2\pi m}{\log q}\log\left|\frac{x}{L}\right|\right\},~(m=\pm 1,\pm 2,\cdots).
\end{equation}
The $S_n(x)$ and $Q_m(x)S_n(x)$ satisfy the same eigenvalue equation; and the singular structure disappears from the probability amplitude $|Q_m(x)S_n(x)|^2=|S_n(x)|^2$. Thus, $Q_m(x)S_n(x)$ also satisfies the same boundary condition as $S_n(x)$. This phase ambiguity is due to the invariance of the Hamiltonian $\hat{H}_0$ such as
\begin{equation}
 \hat{H}_0(x,\partial_x)=\hat{H}_0\left(x,\partial_x+i\frac{2\pi m}{\log q}\frac{1}{x}\right),~(0<x<L).
\end{equation}
In other words, we can say that the gauge-potential like term ${\displaystyle \frac{2\pi m}{\log q}\frac{1}{x}}$ affects no physical effect to the q-deformed particle under consideration. This is a peculiar property of such a particle, which appears only for $q\neq 1$.

\section{Summary and Discussion}

In this paper, we have studied the particle associated with the q-deformed phase space, to which the momentum operator in quantized theory is given by $\hat{p}_q=-\frac{i\hbar}{2}(\hat{\partial}_x+\hat{\partial}_x^\dag)$, where $\hat{\partial}_x$ is the operator characterized by the q-deformed Heisenberg algebra $\hat{\partial}_x x-qx\hat{\partial}_x =1$. Our approach to this q-deformed phase space is to represent the deformed operator $\hat{\partial}_x$ as one in ordinary space by means of a mapping from operators $(x,\partial_x)$; in other words, we tried to represent the q-deformed dynamics as an effective theory in ordinary variable spaces. 

In  section 2, the discussion is made on the free particle in the q-deformed phase space without boundary. We first derived effective action for such a particle represented in the ordinary variable spaces using the path integral method. The form of the action is evaluated from the short time propagator, though it is limited in the application. This is due to the reason that the q-exponential function giving the plane wave in the deformed space does not satisfy the associative law in the usual sense.  If we apply this action rather to a long-time motion, however, we can derive an interesting result such that the trajectories of the particle draw as if they are bounded in configuration space. Further, we evaluated the exact propagator, and found that the propagation from $x$ to $q^{\pm}x$ in the q-deformed phase space is corresponding to one from $x$ to $x$ in ordinary phase space. We note that for the short-time action, the introduction of external potential is simply an additional effect.

The section 3 is the case of compact space bounded by perfectly rigid walls placed in $x=0$ and $x=L$; that is, we have discussed the particle in the box $0<x<L$. Then, we can solve the eigenvalue problem using a pair of $q$-$\sin$ functions such that one of those are independent functions chosen by the boundary conditions $\bar{S}_n(0)=\bar{S}(L)_n=0$, and the others are characterized by the eigenvalue equation $D_x^2S_n=-k_n^2S_n$. Those two $\sin$ functions tend to the standard $\sin$ function as $q\rightarrow1$. By virtue of those sin functions, the energy eigenvalues can be determined as $E_n=\frac{\hbar^2}{2m_q}\left(\frac{\pi_q(n)}{L}\right)^2,(n=1,2,\cdots)$, where $\pi_q(n)~(\rightarrow \pi n,~q\rightarrow 1) $ is a q-dependent function of $n$. The $q$ dependence of $\pi_q(n)$ is fairly large; and, it becomes a rapid increase function of $n$ for a large $q$. 

The boundary condition is not limited to the above case; and, we can consider the case $\psi(0)=1$ and $\psi(\frac{L}{2})=0$. In this case, we need a pair of $q$-$\cos$ functions $\cos_q(x)$ and ${\rm \bar{c}os}_q(x)$ to construct a space of wave functions belonging to the eigenspace of $D_x^2$ followed by the boundary conditions. By definition (\ref{q-bar-cos}), the ${\rm \bar{c}os}_q(x)$ has the product form
\begin{equation}
 {\rm\bar{c}os}_q(x) =\frac{1}{2}\lim_{N\rightarrow \infty}\left[\left(1\dot{+}\frac{ix}{[N]}\right)_q^N+\left(1\dot{-}\frac{ix}{[N]}\right)_q^N \right] \propto \left[ \prod_{k=0}^{N-1}e^{i\theta_k^{(N)}}+1 \right] ,
\end{equation}
where $\theta_k^{(N)}$ is given by (\ref{theta-k}). Thus, for a zero of ${\rm\bar{c}os}_q(x)$, the $\theta_k^{(N)}(x)$ satisfy $\sum_{k=0}^{N-1}\theta_k^{(N)}(x)=(2n+1)\pi$ with an integral $n$. From these conditions, one can find the zeros of ${\rm \bar{c}os}_q(x)$ similarly as in the case of ${\rm \bar{s}in}(x)$ at $x=\pi_q(n+\frac{1}{2}),~(n=0,\pm 1,\pm 2,\cdots)$. Under the normalization ${\rm \bar{c}os}_q(0)=1$, thus, we can obtain the expression
\begin{equation}
 {\rm \bar{c}os}_q(x)=\prod_{n=1}^\infty \left[1-\left(\frac{x}{\pi_q(n-\frac{1}{2})} \right)^2 \right] .
\end{equation}
The ${\rm\bar{c}os}_q(x)$ is an even function obviously; however, its odd property between $x=0$ and $x=L$ is lost because of ${\rm \bar{c}os}_q(\pi)\neq -1$. 

The results obtained in section 3. gives us an insight for the mass  spectrum of a particle embedded in a higher-dimensional spacetime with this type of compact spaces as the extra dimensions. For example, the Kline-Gordon equation in five-dimensional spacetime with q-deformed fifth dimension~\cite{PTP-N-T}
\begin{equation}
 (\partial_\mu \partial^\mu-D_5^2 + m^2)\psi(x^\mu,x^5)=0
\end{equation}
leads to a mass spectrum such that only few states are corresponding to light-mass particles by adjusting the parameter $q$ suitably. These problems will be discussed elsewhere.

\section*{Acknowledgements}
The authors wish to thank the members of the theoretical group in Nihon University for their interest in this work and comments. Fruitful discussion with Mr. R. Asuma, an early collaborator in this work, is also acknowledged.

\appendix
\section{On the eigenstates of $\hat{p}_q$ in a free boundary space }

The key to study the eigenstates of $\hat{p}_q=-i\hbar\left(\frac{q+1}{2q}\right)D_x$ is the equation
\begin{equation}
 D_xx^n=[n]x^{n-1},~(n =0,1,2,\cdots). \label{q-derivative-1}
\end{equation}
From this, one can verify that the q-exponential function defined by~\cite{EXTON}
\begin{equation}
 e_q(x)=e_q^x=\sum_{n=0}^\infty \frac{1}{[n]!}x^n,~(~[0]! \equiv 1~{\rm and}~[n]!=\prod_{k=1}^n[k]~{\rm for}~n\geq 1)
\end{equation}
satisfies
\begin{equation}
 D_xe_q(ax)=ae_q(ax)~,~(~a=const.~). \label{q-derivative-2}
\end{equation}
It is obvious by definition that $e_q(x) \rightarrow e^x,~(q \rightarrow 1)$ and $e_q(x)e_q(y) \neq e_q(x+y)$ for $q\neq 1$. Since $[n]\geq n$, the series of $e_q(x)$ is rapidly convergent than the usual exponential function. In particular, $e_q(x)\rightarrow 0$ according as $x\rightarrow -\infty$. Using this q-exponential function we can introduce the q-analog of $e^{ix}=\cos(x)+i\sin(x)$ so that
\begin{equation}
 e_q(ix)=\cos_q(x)+i\sin_q(x)~,
\end{equation}
from which we can write explicitly
\begin{align}
 \cos_q(x) &=\frac{1}{2}\left(e_q(ix)+e_q(-ix)\right)=\sum_{k=0}^\infty \frac{(-1)^k}{[2k]!}x^{2k}, \\
 \sin_q(x) &=\frac{1}{2i}\left(e_q(ix)-e_q(-ix)\right)=\sum_{k=0}^\infty \frac{(-1)^k}{[2k+1]!}x^{2k+1}.
\end{align}

Now the function
\begin{equation}
 \langle x|k,q\rangle=N_qe_q(ikx)~,~(N_q=const.)
\end{equation}
is an eigenstate of $\hat{p}_q$ belonging to the eigenvalue $\hbar k$, though it is not a simple plane-wave function in the ordinary $x$ space. Here, the wave number $k$ may be an ordinary continuous real number running from $-\infty$ to $\infty$.

To determine the constant $N_q$, let us evaluate the inner product $\langle k,q|k^\prime, q\rangle$ by assuming $|k|<|k^\prime|$ without loss of generality. Then, introducing a factor $\epsilon(\rightarrow +0)$ to define a convergent inner product, we have
\begin{align}
 \langle k,q|k^\prime, q\rangle &= |N_q|^2 \int_{-\infty}^\infty dxe_q^{-ikx-\frac{1}{2}\epsilon|x|}e_q^{ik^\prime x-\frac{1}{2}\epsilon|x|} \nonumber \\
 &=|N_q|^2\sum_{n=0}^\infty \left\{-i\left(k-\frac{i\epsilon}{2}\right) \right\}^n \frac{1}{[n]!}\int_{0}^\infty dx x^n e_q^{i(k^\prime + \frac{i\epsilon}{2})x}+ c.c. \nonumber \\
 &=|N_q|^2\sum_{n=0}^\infty\frac{i}{k^\prime+\frac{i\epsilon}{2}}\left(\frac{k-\frac{i\epsilon}{2}}{k^\prime+\frac{i\epsilon}{2}}\right)^n\frac{1}{[n]!}\int_0^\infty dte_q^{-t}t^n +c.c.~, \label{inner-product}
\end{align}
where the analytic continuation $i(k^\prime+\frac{i\epsilon}{2})x \rightarrow -t$ has been done. Further, due to the scale invariance of $dt/\{(q-q^{-1})t\}$, it holds that $\int_0^\infty dt \{D_tf(t)\}g(t)=-\int_0^\infty dtf(t)\{D_tg(t)\}$ for a convergent integral. Thus, by taking (\ref{q-derivative-1}) and (\ref{q-derivative-2}) into account, we obtain
\begin{equation}
 \int_0^\infty dte_q^{-t}t^n=-\int_0^\infty dt\{D_te_q^{-t}\}t^n=[n]\int_0^\infty dte_q^{-t}t^{n-1}=\cdots=[n]!\Gamma_q[1] \label{q-gamma-1}
\end{equation}
with the definition of q-gamma function
\begin{equation}
 \Gamma_q[n]= \int_0^\infty dt e_q^{-t}t^{n-1}~,~(n=1,2,\cdots)~, \label{q-gamma-2}
\end{equation}
to which one can see that $\Gamma_q[1]\neq 1$. Therefore, substituting (\ref{q-gamma-1}) for (\ref{inner-product}), we arive at the expression
\begin{align}
 \langle k,q|k^\prime, q\rangle &= |N_q|^2 \left( \frac{i}{k^\prime-k+i\epsilon }+c.c.\right) \nonumber \\
 &=2\pi|N_q|^2 \times \frac{1}{\pi}\frac{\epsilon}{(k^\prime-k)^2+\epsilon^2} \nonumber \\
 &=\delta(k^\prime-k)
\end{align}
on condition that 
\begin{equation}
 N_q=\frac{1}{\sqrt{2\pi\Gamma_q[1]}}~.
\end{equation}

For applying $\langle x|k,q\rangle$ to the propagation kernel, it is worthwhile to study in more detail the product $e_q(x)e_q(y)$, which can be written as
\begin{equation}
 e_q(x)e_q(y)=\sum_{N=0}^\infty \frac{1}{[N]!}(x\dot{+}y)_q^N~, \label{q-exponential-product}
\end{equation}
where $(x\dot{+}y)_q^N$ is the q-binomial expansion defined by
\begin{equation}
  (x\dot{+}y)_q^N=\sum_{n+m=N}\frac{[N]!}{[n]![m]!}x^n y^m~. \label{q-binomial}
\end{equation}
We have used the symbol \lq\lq$\dot{+}$\rq\rq to stress that $(x\dot{+}y)_q^N$ is not a function of $x+y$. Now, the q-binomial expansion satisfies the following recursion formula:
\begin{equation}
 (x\dot{+}y)_q^{N+1}=(x+q^{N}y)(x\dot{+}q^{-1}y)_q^{N}~. \label{recursion}
\end{equation}
Indeed, it is not difficult to verify that
\begin{align}
 x(x\dot{+}q^{-1}y)_q^{N} &=\sum_{n=0}^N\frac{[N]!}{[n]![N-n]!}x^{n+1}y^{N-n}q^{-N+n}~,~(k=n+1) \nonumber \\
 &=\sum_{k=1}^{N+1}\frac{[N+1]!}{[k]![(N+1)-k]!}x^ky^{(N+1)-k}\times \frac{[k]q^{-(N+1)+k}}{[N+1]} \label{recursion-1}
\end{align}
and
\begin{align}
 q^Ny(x\dot{+}q^{-1}y)_q^{N} &=\sum_{n=0}^N\frac{[N]!}{[n]![N-n]!}x^{n}y^{(N+1)-n}q^{n}~,~(k=n) \nonumber \\
  &=\sum_{k=1}^{N+1}\frac{[N+1]!}{[k]![(N+1)-k]!}x^ky^{(N+1)-k}\times \frac{[(N+1)-k]q^{k}}{[N+1]}~. \label{recursion-2}
\end{align}
Here, $\left.\frac{[k]q^{-(N+1)+k}}{[N+1]}\right|_{k=N+1}=1$, $\left.\frac{[(N+1)-k]q^{k}}{[N+1]}\right|_{k=0}=1$ and $\frac{[k]q^{-(N+1)+k}}{[N+1]}+\frac{[(N+1)-k]q^{k}}{[N+1]}=1$ for $1\leq k\leq N$; thus, the sum of equations (\ref{recursion-1}) and (\ref{recursion-2}) becomes the left-hand side of Eq.(\ref{recursion}). Using Eq.(\ref{recursion}) repeatedly, we can also obtain a factorized form to the q-binomial expansion:
\begin{align} 
 (x\dot{+}y)_q^N &=(x+q^{-N+1}y)(x+q^{-N+3}y)\cdots(x+q^{N-1}y) \nonumber \\
 &=\prod_{k=0}^{N-1}(x+q^{N-1-2k}y). \label{product-formula}
\end{align}
If it is necessary, $e_q^xe_q^{-y}$, $(x\dot{-}y)_q^N$ and its factorized form are obtained by substituting $-y$ for $y$ in Eqs.(\ref{q-exponential-product}), (\ref{q-binomial}) and (\ref{product-formula}) respectively. Then, it can be found that $(x\dot{-}y)_q^N$ has simple zeros at $x=q^{N-1-2k}y,~(k=0,1,\cdots N-1)$ in contrast that $(x-y)^N$ has zero of order $N$ at $x=y$. In particular, since $(x\dot{-}y)^{2k},(k=1,2,\cdots)$ have zeros at $y=q^{\pm}x$, we have
\begin{align}
 Re\left\{e_q(ix)e_q(-iq^{\pm 1}x)\right\} &=\cos_q(x)\cos_q(q^{\pm 1}x)+\sin_q(x)\sin_q(q^{\pm 1}x) \nonumber \\
 &=\sum_{k=0}^\infty \frac{(-1)^k}{[2k]!}(x\dot{-}q^{\pm 1}x)^{2k}=1~.
\end{align} 

\section{Eigenvalue problem of $\hat{p}_q^2$ for a particle in a box}

The eigenvalue problem for a particle in the box $0\leq x\leq L$ shows fairly different aspect from the $q=1$ case. To study the problem, we introduce another type of q-exponential function defined by~\cite{EXTON,Slater}
\begin{equation}
 \bar{e}_q(x)=\sum_{n=0}^\infty \frac{x^n}{[n]!}q^{-\frac{1}{2}n(n-1)}=\sum_{n=0}^\infty \frac{x^n}{[n,q^2]!},
\end{equation}
where
\begin{equation}
 [n,q^2]=\frac{q^{2n}-1}{q^2-1}~.
\end{equation}
The importance of this function is that with help of $\frac{[N]!}{[r]![N-r]!}\sim \frac{[N]^r}{[r]!}q^{-\frac{1}{2}r(r-1)}$ for large $N$, one can obtain the expression
\begin{equation}
 \bar{e}_q(x)=\lim_{N\rightarrow\infty}\left(1\dot{+}\frac{x}{[N]}\right)_q^N~. \label{factorization}
\end{equation}
The q-sin/cos functions in this case can be defined, as usual, by
\begin{align}
 {\rm\bar{s}in}_q(x) &=\frac{1}{2i}\left(\bar{e}_q(ix)-\bar{e}_q(-ix)\right)~,\\
 {\rm\bar{c}os}_q(x) &=\frac{1}{2}\left(\bar{e}_q(ix)+\bar{e}_q(-ix)\right) . \label{q-bar-cos}
\end{align}
It should be noticed that the $e_q(x)$ and $\bar{e}_q(x)$ are eigenfunctions of the difference operators $D_x\equiv x^{-1}[\hat{N}]$ and $\bar{D}_x\equiv x^{-1}[\hat{N},q^2]$ respectively; that is, we have
\begin{equation}
 D_x e_q(ax)=ae_q(ax), \hspace{5mm} \bar{D}_x \bar{e}_q(ax)=a\bar{e}_q(ax).
\end{equation}
However, we can also verify that
\begin{equation}
 D_x \bar{e}_q(ax)=a\bar{e}_q(aq^{-1}x), \hspace{5mm} \bar{D}_x e_q(ax)=ae_q(aqx).
\end{equation}

Now, the characteristics of $\bar{e}_q(x)$ is that the ${\rm\bar{s}in}(x)={\rm Im}\bar{e}_q(ix)$ has an infinite product representation, an analogue of $\sin(x)=x\prod_{n=1}^\infty\left(1-\frac{x^2}{(\pi n)^2}\right)$ for $q=1$, with the aid of (\ref{factorization}). To derive the formula, we apply the q-binomial expansion (\ref{product-formula}) to (\ref{factorization}). Then, we obtain
\begin{align}
 {\rm\bar{s}in}_q(x) &=\frac{1}{2i}\lim_{N\rightarrow \infty}\left[\left(1\dot{+}\frac{ix}{[N]}\right)_q^N-\left(1\dot{-}\frac{ix}{[N]}\right)_q^N \right] \nonumber \\
 &=\frac{1}{2i}\lim_{N\rightarrow \infty}\prod_{k=0}^{N-1}\left(1-q^{N-1-2k}\frac{ix}{[N]}\right)\times\left[\prod_{k=0}^{N-1}\left(\frac{1+q^{N-1-2k}\frac{ix}{[N]}}{1-q^{N-1-2k}\frac{ix}{[N]}}\right)-1\right]. \label{bar-sin}
\end{align}
It is obvious that the zeros of ${\rm\bar{s}in}_q(x)$ exist in the factor $\left[\prod_{k=0}^{N-1}e^{i\theta^{(N)}_k(x)}-1\right]$ in the right-hand side of (\ref{bar-sin}). Here, we have put
\begin{equation}
 e^{i\theta^{(N)}_k(x)}\equiv \frac{1+q^{N-1-2k}\frac{ix}{[N]}}{1-q^{N-1-2k}\frac{ix}{[N]}} ~. \label{theta-k}
\end{equation}
This implies that if $x$ is a zero of ${\rm\bar{s}in}_q(x)$, then the $\theta^{(N)}_k(x)$'s satisfy
\begin{equation}
 \sum_{k=0}^{N-1}\theta^{(N)}_k(x)=2\pi n~,~~(~n=0,\pm 1,\pm 2,\cdots). \label{sum-theta}
\end{equation}
Further, since the $\theta^{(N)}_k(x)$ in equation (\ref{theta-k}) can be solved with respect to $x$ as
\begin{equation}
 x=q^{-N+1+2k}[N]\tan\frac{\theta^{(N)}_k(x)}{2}, \label{zeros}
\end{equation}
we can get the expression
\begin{equation}
 \Theta^{(N)}(x)\equiv \frac{1}{2}\sum_{k=0}^{N-1} \theta^{(N)}_k(x) =\sum_{k=0}^{N-1}\left[\tan^{-1}\left\{\frac{q^{N-1}}{[N]}q^{-2k}x \right\}+\delta \right] . \label{sum-arctan}
\end{equation} 
Here $\delta=\pi \times ({\rm integer})$ is an undetermined phase coming from the periodicity of $\tan(x)$, which can be absorbed by the right-hand side of equation (\ref{sum-arctan}); and so, we put hereafter $\delta=0$.

This equation tells us the structure of zeros of ${\rm \bar{s}in}_q(x)$. First, it is obvious that $\Theta^{(N)}(x)$ tends to $x$ as $q\rightarrow 1$ because of $q^{N-1}/[N]\sim N^{-1}$ for $q\sim 1$. Namely, the zeros of ${\rm\bar{s}in}_q(x)$ become $\pi n,(n=0,\pm 1,\cdots)$ as expected in this limit. 

Secondly, for $q \neq 1$,  we obtain the asymptotic behavior $\Theta^{(N)}(x)\sim \frac{N \pi}{2}-\frac{[N]^2}{x}$ for a large $x$ by taking $\tan^{-1}(ax)\sim \frac{\pi}{2}-\frac{1}{ax}$ into account. Then, $\Theta^{(N)}(x)$ tends to the upper bound $\frac{\pi N}{2}$ as $x\rightarrow \infty$ for a fixed $N$; and so, the number of zeros determined by $\Theta^{(N)}(x)=\pi n$ becomes finite; in other words, the $\Theta^{(N)}(x)$ does not have its inverse function. Since, however, $\Theta^{(N)}(x)$ is an increase function of $N$, ${\displaystyle y=\Theta(x)\equiv\lim_{N\rightarrow \infty}\Theta^{(N)}(x)}$ can be solved with respect to $x$. To do this, we write
\begin{equation}
\Theta(x)=\sum_{k=1}^\infty a_k x^k , \label{Theta-infty}
\end{equation}
where $a_2=a_4=\cdots=0$ and
\begin{equation}
 a_{2k-1}=\frac{(-1)^{k-1}}{2k-1}\frac{(1-q^{-2})^{2k-1}}{1-q^{-2(2k-1)}},~(k=1,2,\cdots).
\end{equation}
Since the $\Theta(x)$ is a single valued analytic function of $x$, $y=\Theta(x)$ can be inverted by the Lagrange expansion
\footnote{The $b_n$'s can be calculated by\cite{Lagrange-expansion}
$b_n=\frac{1}{n!}\left[ \left(\frac{d}{dx}\right)^{n-1} \left(\frac{x}{\Theta(x)} \right)^n \right]_{x=0},~(n=1,2,\cdots)$. Writing, here, $\Theta^{\infty}(x)/x=a_1(1+B(x))$ by using $B(x)=\tilde{a}_2x^2+\tilde{a}_4x^4+\tilde{a}_6x^6+\cdots$ with $\tilde{a}_k=a_{k+1}/a_1$, we obtain the expression $ b_n=\frac{1}{a_1^n n!}\left[\left(\frac{d}{dx}\right)^{n-1}\sum_{k=0}^\infty \begin{pmatrix} -n \\ k\end{pmatrix}B(x)^k\right]_{x=0}$, which leads to for $n=2m+1$
\[ b_{2m+1}=\frac{1}{(2m+1)a_1^{2m+1}}\sum_{k=1}^m \begin{pmatrix} -(2m+1) \\ k \end{pmatrix}\sum_{i_1+\cdots+i_k=2m}\tilde{a}_{i_1}\cdots\tilde{a}_{i_k}, (m=1,2,\cdots). \]
}  
 $x=\sum_{k=1}^\infty b_k y^k$ with $b_2=b_4=\cdots=0$ and
\begin{align}
b_1 &= \frac{1}{a_1}=1, \label{b-1} \\
b_3 &= -\frac{a_3}{a_1^4}=\frac{1}{3}\frac{(1-q^{-2})^3}{1-q^{-6}}, \\
b_5 &= -\frac{a_5}{a_1}+3\frac{a_3^2}{a_1^7}=-\frac{1}{5}\frac{(1-q^{-2})^5}{1-q^{-10}}+\frac{1}{3}\left\{\frac{(1-q^{-2})^3}{1-q^{-6}}\right\}^2, \\
b_7 &=-\frac{a_7}{a_1^8}+4\frac{a_3a_5}{a_1^9}-12\frac{a_3^3}{a_1^{10}} \nonumber \label{b-7} \\
 &=\frac{1}{7}\frac{(1-q^{-2})^7}{1-q^{-14}}-\frac{8}{15}\frac{(1-q^{-2})^8}{(1-q^{-6})(1-q^{-10}))}+\frac{4}{9}\frac{(1-q^{-2})^9}{(1-q^{-6})^3} ,
\end{align}
and so on. Therefore, writing $\pi_q(n)=\sum_{k=1}^\infty b_k(\pi n)^k$ and by taking into account that $\Theta(x)$ is an odd function of $x$, we obtain the expression
\begin{equation}
 {\rm\bar{s}in}_q(x)=x\prod_{n=0}^\infty\left[1-\left( \frac{x}{\pi_q(n)} \right)^2 \right] \label{bar-sin}
\end{equation}
under the normalization ${\rm\bar{s}in}_q(x)/x\rightarrow 1,(x\rightarrow 0)$. Since $\pi_q(n)$ tends obviously to $\pi n$ as $q\rightarrow 1$, the right-hand side of (\ref{bar-sin}) reduces to the standard product formula of $\sin(x)$ in this limit. 

\begin{figure}
\centering
\includegraphics[width=6cm]{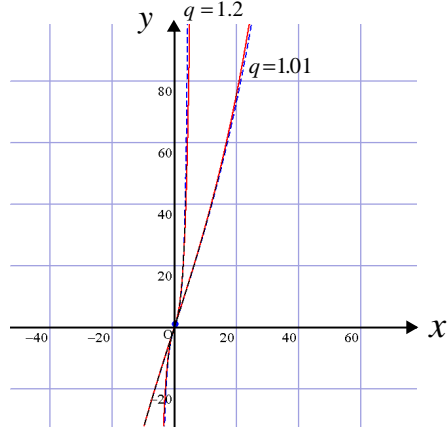}
\caption{The real lines are numerical curves of $\pi_q(x)$ for $q=1.01$ and $q=1.2$, respectively. On the other hands, the dotted lines show the counterparts  obtained by the Lagrange expansion up to the order of $x^7$. The both types of lines show good agreement for small value of $x$ as expected. For a larger $q$, the property of stationary curves will be revealed to the dotted lines. In particular, the absolute values of zero points become rapidly increase, according to $q$ increases.}
\end{figure}

Now, with aid of the product formula (\ref{bar-sin}), one can find that the function
\begin{equation}
 \bar{S}_{n}(x)\equiv N_n{\rm\bar{s}in}_q(k_n x),~(~N_n=const.,~k_n=\pi_q(n)/L~)
\end{equation}
is a solution of the eigenvalue problem
\begin{equation}
 \bar{D}^2 \bar{S}_n(x)=-k_n^2 \bar{S}_n(x)~,~~(n=1,2,\cdots)
\end{equation}
under the boundary conditions $\bar{S}_n(0)=\bar{S}_n(L)=0$. 

Turning back to the eigenvalue problem of $D^2$, let us define
\begin{equation}
 S_n(x)\equiv N_n\sin_q(x)=q^{\frac{1}{2}\hat{N}(\hat{N}-1)}\bar{S}_n(x).
\end{equation}
Then taking $D^2=q^{\frac{1}{2}\hat{N}(\hat{N}-1)}\bar{D}^2q^{-\frac{1}{2}\hat{N}(\hat{N}-1)}$ into account, we obtain
\begin{equation}
 (D^2+k_n^2)S_n(x)=q^{\frac{1}{2}\hat{N}(\hat{N}-1)}\left\{(\bar{D}^2+k_n^2)\bar{S}_n(x)\right\}=0.
\end{equation}
Thus, the function $S_n(x)$ is also an eigenstate of $D^2$ belonging to the eigenvalue $-k_n^2$, which is determined by the zero point of the function $\bar{S}_n(x)$ instead of $S_n(x)$. 

The normalization of $\{ S_n(x)\}$ can be done by introducing the q-integral defined by the inverse $D^{-1}=[\hat{N}]^{-1}x=x[\hat{N}+1]^{-1}$ in such a way that
\begin{equation}
 \int_a^b \Delta_qf(x)\equiv \left. D^{-1}f(x)\right|_a^b=\left. \sum_{k=0}^\infty \Delta_q x q^{-(2k+1)}f(q^{-(2k+1)}x)\right|_a^b~,~(q>1), \label{q-integral}
\end{equation}
where $\Delta_q x=(q-q^{-1})x$. The right-hand side of (\ref{q-integral}) reduces to the standard Riemann integral of $f(x)$ according as $q\rightarrow 1$; one can also verify that 
\begin{equation}
 \int_a^b \Delta_qx \left\{ Df(x)\right\}=f(x_b)-f(x_a)
\end{equation}
provided that the infinite series of (\ref{q-integral}) converge. Then, integrating the both sides of the equation
\begin{equation}
 D\left[S_n(qx)DS_m(x)\right]=DS_n(qx)DS_m(qx)+S_n(x)D^2S_m(x),
\end{equation}
from $0$ to $x_0$, we obtain
\begin{equation}
 \left[S_{n}(qx)DS_{m}(x)-S_{m}(qx)DS_{n}(x)\right]_{0}^{x_0}=-(k_m^2-k_n^2)\langle S_nS_m \rangle~, \label{orthogonality}
\end{equation}
where $\langle( \cdots )\rangle=\int_0^{x_0}\Delta_q x(\cdots)$ and $x_0$ is a point characterized by $S_n(qx_0)=0,~(n=1,2,\cdots)$. Thus, $S_n$'s satisfy the orthogonality $\langle S_nS_m \rangle=\delta_{n,m}$ by adjusting the normalization constants $N_n$'s appropriately. We, here, suppose that $\{S_n(x)\}$ forms a complete basis of $\{\bar{S}_n(x)\}$ space. Then, we have
\begin{equation}
 \bar{S}_n(x)=q^{-\frac{1}{2}\hat{N}(\hat{N}-1)}S_n(x)=\sum_{m=1}^\infty S_m(x)\langle S_m q^{-\frac{1}{2}\hat{N}(\hat{N}-1)}S_n\rangle , \hspace{3mm} \label{complete-1}
\end{equation}
from which follows
\begin{align}
 S_n(x) &= q^{\frac{1}{2}\hat{N}(\hat{N}-1)}\bar{S}_n(x)=q^{\frac{1}{2}\hat{N}(\hat{N}-1)}\sum_{m=1}^\infty S_m(x)\langle S_m q^{-\frac{1}{2}\hat{N}(\hat{N}-1)}S_n\rangle \nonumber \\
 &=\sum_{m=1}^\infty \bar{S}_m(x)\langle S_m q^{-\frac{1}{2}\hat{N}(\hat{N}-1)}S_n\rangle . \label{complete-2}
\end{align}
The expansion (\ref{complete-2}) means that $S_n(x)$ satisfies the boundary conditions $S_n(L)=0$ in addition to $S_n(0)=0$. On the other hand, the expansion (\ref{complete-1}) requires $\bar{S}_n(qx_0)=0$ to ensure the orthogonality of $\{S_n(x)\}$. Therefore, by choosing $x_0=q^{-1}L$, both expansions (\ref{complete-1}) and (\ref{complete-2}) become consistent.

\end{document}